\documentclass[12pt]{iopart}
\usepackage{iopams}
\usepackage{graphicx}
\usepackage{setstack}
\usepackage{enumerate}
\usepackage{algorithm}
\usepackage{algpseudocode}
\let\leftroot\relax
\let\uproot\relax

\usepackage{mmacells}
\usepackage{appendix}
\usepackage{relsize}
\usepackage[T1]{fontenc}

\newcounter{defi}[section]
\renewcommand{\thedefi}{\thesection.\arabic{defi}}

\newenvironment{defi}[1][]{
  \refstepcounter{defi}
  \par\noindent\textbf{Definition~\thedefi.} {(#1)}\nopagebreak}{\par
}

\newcounter{prop}[section]
\renewcommand{\theprop}{\thesection.\arabic{prop}}

\newenvironment{prop}[1][]{
  \refstepcounter{prop}
  \par\noindent\textbf{Proposition~\theprop.} \nopagebreak}{\par
}

\newcounter{rem}[section]
\renewcommand{\therem}{\thesection.\arabic{rem}}

\newenvironment{rem}[1][]{
  \refstepcounter{rem}
  \par\noindent\textbf{Remark~\therem.} \nopagebreak}{\par
}

\newcounter{thm}[section]
\renewcommand{\thethm}{\thesection.\arabic{thm}}

\newenvironment{thm}[1][]{
  \refstepcounter{thm}
  \par\noindent\textbf{Theorem~\thethm.} {(#1)}\nopagebreak}{\par
}

\newenvironment{proof}[1][Proof]{%
  \par\noindent\textit{#1:}\quad
}{%
  \hfill$\square$\par
}

\begin{document}
\title[Effective criteria for entanglement witnesses]{Effective criteria for entanglement witnesses in small dimensions}
\author{Łukasz Grzelka\textsuperscript{1}, Łukasz Skowronek\textsuperscript{1} and
Karol Życzkowski\textsuperscript{1,2}}
\address{\textsuperscript{1}Institute of Theoretical Physics, Jagiellonian University, Kraków, Poland\\
\textsuperscript{2}Center for Theoretical Physics, Polish Academy of Sciences, Warsaw}

\ead{lukasz.grzelka@doctoral.uj.edu.pl; 
    lukasz.m.skowronek@gmail.com;
    karol.zyczkowski@uj.edu.pl}
\begin{abstract}
We present an effective set of necessary and sufficient criteria for block-positivity of matrices of order $4$ over $\mathbb{C}$. The approach is based on Sturm sequences and quartic polynomial positivity conditions presented in recent literature. The procedure allows us to test whether a given $4\times 4$ complex matrix corresponds to an entanglement witness, and it is exact when the matrix coefficients belong to the rationals, extended by $\mathrm{i}$. The method can be generalized to $\mathcal{H}_2\otimes\mathcal{H}_d$ systems for $d>2$ to provide necessary but not sufficient criterion for block-positivity. We also outline an alternative approach to the problem relying on Gröbner bases.
\end{abstract}
\noindent{\it Keywords}: Entanglement witness, Quantum Entanglement, Block-Positivity.
\noindent{}
\maketitle
\section{Introduction}
A~fundamental feature of composite quantum systems is the effect of entanglement, a property which in quantum theory often appears as a resource required to perform a given task such as quantum teleportation, superdense coding or quantum key distribution \cite{Nielsen_Chuang, RevModPhys.81.1301}. In this work, we analyze operators called \textit{entanglement witnesses}, suitable for detecting entangled states. The construction of these operators was proposed in~\cite{HORODECKI19961}, while the phrase ``entanglement witnesses'' was introduced by Terhal \cite{TERHAL2000319}. A comprehensive review of the theory behind entanglement witnesses along with their experimental application is provided in \cite{G_hne_2009}. Their classification and construction was also discussed in \cite{Chruściński_2014}. For more information concerning quantum entanglement consult the review \cite{RevModPhys.81.865}. 

A well-known computationally difficult problem in quantum theory is that of distinguishing between entangled and separable mixed states. A related membership problem asks which matrices belong to the set of entanglement witnesses. As it turns out, such elements simultaneously belong to set $\mathcal{B}$, the dual of separable states $\Omega_\mathrm{S}$, and do not belong to the space of states $\Omega$ (see Figure \ref{fig:1} for reference). In this work we present a procedure answering this membership problem for the simplest case of matrices of order 4, and explain how it applies for cases of higher dimension. 
For previous attempts at addressing this issue see  \cite{Skowronek_2009}. More recent papers tackling related problems include \cite{Jannesary2025,pastuszak2024oneparameterfamilieshermiticitypreservingsuperoperators}

This paper is organized as follows. Section \ref{s2} introduces preliminary information on entanglement witnesses and block-positivity. In Section \ref{s3} we present an effective procedure for testing whether a given two-qubit operator is an entanglement witness, while in Section \ref{beyond} we discuss how this procedure provides necessary criteria in systems of higher dimensions. In Section \ref{s5} we present an alternative method relying on Gröbner bases, which provides sufficient but not necessary criteria for entanglement witnesses in qudit-qubit systems. We conclude in Section \ref{s6} by applying the presented methods to two one parameter families of qubit-qubit and qutrit-qubit operators.
\section{Setting the scene}\label{s2}
 Let $\mathcal{L}\left(\mathcal{L}(\mathcal{H}_\mathrm{A}),\mathcal{L}(\mathcal{H}_\mathrm{B})\right)$ denote the space of maps $\Phi:\mathcal{L}(\mathcal{H}_\mathrm{A})\rightarrow\mathcal{L}(\mathcal{H}_\mathrm{B})$ connecting spaces of operators acting on $\mathcal{H}_\mathrm{A}$ and $\mathcal{H}_\mathrm{B}$ respectively. Such a map $\Phi$ is called \textit{positive} if it preserves positive operators i.e. $\Phi(A)\geq 0\;\forall A\geq0$. States in a quantum system are represented by density matrices $\Omega\ni\rho\geq0$, and hence any physical operation $\Phi$ has to be positive. Additionally, a physically implementable operation should remain positive for all extensions
\begin{eqnarray}\label{eq:CP}
    \Phi\otimes\mathbb{I}_k: \mathcal{L}(\mathcal{H}_\mathrm{A}\otimes\mathcal{H}_k)\rightarrow\mathcal{L}(\mathcal{H}_\mathrm{B}\otimes\mathcal{H}_k)
\end{eqnarray}
by an identity operator $\mathbb{I}_k$ of an arbitrary dimension $k$. A map $\Phi$ satisfying criterion (\ref{eq:CP}) is called \textit{completely positive} \cite{Choi}.

The space of maps $\mathcal{L}\left(\mathcal{L}(\mathcal{H}_\mathrm{A}),\mathcal{L}(\mathcal{H}_\mathrm{B})\right)$ is isomorphic to the space $\mathcal{L}\left(\mathcal{H}_\mathrm{AB}\right)$ of operators acting on the bipartite system $\mathcal{H}_\mathrm{AB}=\mathcal{H}_\mathrm{A} \otimes \mathcal{H}_\mathrm{B}$, via \textit{Choi-Jamiołkowski isomorphism} \cite{Chruściński_2014,Skowronek_2010}. That is, a map $\Phi: \mathcal{L}(\mathcal{H}_\mathrm{A})\rightarrow\mathcal{L}(\mathcal{H}_\mathrm{B})$ can be associated with a so-called Choi matrix $C_\Phi = \left(\mathbb{I}_\mathrm{A}\otimes\Phi\right)|\phi^+\rangle\langle\phi^+|\in\mathcal{L}(\mathcal{H}_{\mathrm{AB}})$, where $|\phi^+\rangle$ is a maximally entangled state in $\mathcal{H}_\mathrm{A}\otimes\mathcal{H}_\mathrm{A}$. Of particular interest are  Choi matrices corresponding to positive and completely positive maps. According to Choi and Jamiołkowski theorems \cite{Choi, Jamiołkowski_BP}, the former correspond to positive semi-definite matrices, while the latter correspond to matrices satisfying the following \textit{block-positivity} property.
\medskip
\begin{defi}[Block-positivity]\label{defi:Block_Positivity}
Let $A$ be an operator acting on bipartite system $\mathcal{H}_\mathrm{AB}$,  of the form $\mathcal{H}_\mathrm{A}\otimes\mathcal{H}_\mathrm{B}$. An operator $A$ is block-positive if it satisfies
\begin{eqnarray}\label{eq:bp1}
    \forall |v\rangle\in\mathcal{H}_\mathrm{A} \land \forall |w\rangle\in\mathcal{H}_\mathrm{B},\quad \langle v\otimes w|A|v\otimes w\rangle \geq 0.
\end{eqnarray}
The set of block-positive operators is denoted as $\mathcal{B}$.
\end{defi}

\begin{figure}[h]
    \centering
    \includegraphics[width=0.6\linewidth]{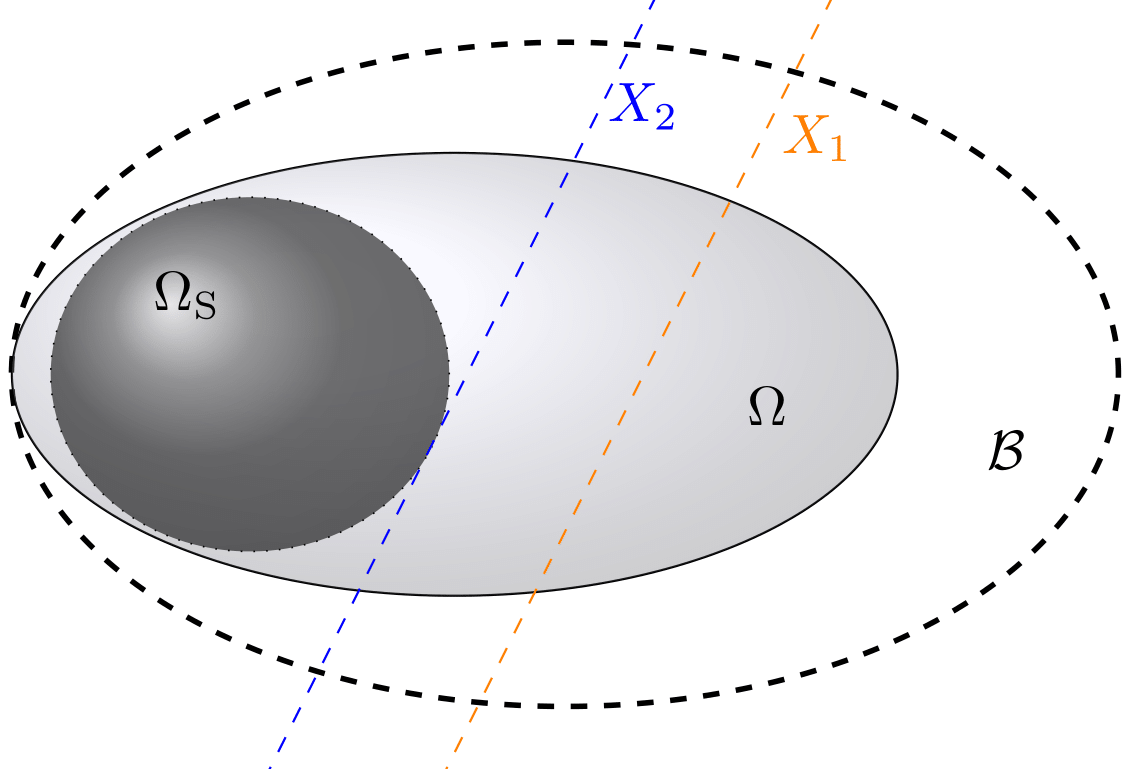}
    \caption{Diagram representing the set of density matrices $\Omega$, with the subset of separable states $\Omega_\mathrm{S}$. Set $\mathcal{B}$ containing block-positive matrices, is the dual of $\Omega_\textrm{S}$. Dashed lines $X_1$ and $X_2$ represent entanglement witnesses separating $\Omega_\textrm{S}$ from detected entangled states. Line $X_2$ corresponds to an optimal witness and is tangent to the set of separable states. }
    \label{fig:1}
\end{figure}

In the context of entanglement detection, positive but not completely positive maps provide criteria for separability. As noted in the famous witness lemma \cite{HORODECKI19961}, the same is true for the corresponding block-positive matrices. For any entangled state $\rho \in \Omega_\mathrm{AB}$ there exists a block-positive operator $X$ such that $\tr(X\rho)<0$, and $\tr(X\sigma)\geq0$ for any separable $\sigma\in\Omega_\mathrm{AB}$.
Such block-positive but not positive semi-definite operators are called \textit{entanglement witnesses}. An operator $X$ acting on a bipartite system $\mathcal{H}_\mathrm{AB}$ is an entanglement witness if it is Hermitian,  it satisfies $\tr(X\rho_\mathrm{s})\geq0$ for all separable states $\rho_\mathrm{s}$, and there exists at least one entangled state $\rho_\mathrm{e}$ such that $\tr(X\rho_\mathrm{e})<0$. The sets of block-positive operators $\mathcal{B}$ and density operators $\Omega$ are illustrated in Figure \ref{fig:1}. Set $\mathcal{B}$ is dual to the set of separable states $\Omega_\mathrm{S}$ in the sense that $\forall\sigma\in\Omega_\mathrm{S},X\in\mathcal{B}$, we have $\tr(X\sigma)\geq0$. Of particular interest are members of the boundary of set $\mathcal{B}$ in Figure~\ref{fig:1}, that is, entanglement witnesses for which the minimum expectation value among the separable states,
\begin{eqnarray}\label{eq:mu}
    \mu = \min_{\sigma\in\Omega_\mathrm{S}}(\tr(X\sigma))
\end{eqnarray}
is equal to zero. Witnesses satisfying this condition are sometimes referred to as \textit{weakly optimal} \cite{Badziag_2013}, while \textit{optimal witnesses} \cite{Lewenstein_2000} are members of the set $\mathcal{B}$ that cannot be further improved by subtracting positive semi-definite matrices. The definition of witness optimality has later been refined in \cite{HaKye}. From now on we will refer to $\mu$ from (\ref{eq:mu}) as the \textit{minimal local value}. One can also consider the minimum of $\tr\left(X\sigma\right)$ over all $\sigma$ of maximum Schmidt rank $k$, as has recently been done in \cite{chenCollins}.
\section{Block-positivity and entanglement witnesses for two-qubit system}\label{s3}
 The main problem addressed in this paper is how to test whether a given operator is or is not an entanglement witness. The difficult part of the task is checking the block-positivity property. Following the approach presented in \cite{Skowronek_2009}, let $X_{ij,i'j'}$ denote the elements of an operator acting on a bipartite system $\mathcal{H}_\mathrm{A}\otimes\mathcal{H}_\mathrm{B}$ and let $v_i$, $w_j$ denote the elements of $|v\rangle\in\mathcal{H}_\mathrm{A}$ and $|w\rangle\in\mathcal{H}_\mathrm{B}$ respectively. Block-positivity condition (\ref{eq:bp1}) in terms of these elements becomes
 \begin{eqnarray}\label{eq:Elementwise Block-Positivity}
     X_{ij,i'j'}\overline{v}_i\overline{w}_jv_{i'}w_{j'}\geq 0,\;\;\forall v_i,w_j\in\mathbb{C},
 \end{eqnarray}
 where, using the Einstein convention, we have skipped the sum sign for better readability. One can now define a projected operator 
 \begin{eqnarray}\label{eq:defXw}
    (X_w)_{ii'} = X_{ij,i'j'}\overline{w}_jw_{j'},
\end{eqnarray}
and say that an operator $X$ is block-positive if and only if the corresponding projected operator $X_w$ is semi-positive definite for all $|w\rangle \in \mathcal{H}_\mathrm{B}$.

Semi-positivity condition $X_w\geq0$ is equivalent to the non-negativity of sums of principal minors of the projected operator $X_w$. 
In the case of two qubit system, a projected operator $X_w$ is a two by two matrix and $X_w\geq0$ holds if and only if $\tr{X_w}\geq0$ and $\det{X_w}\geq 0$ for all $|w\rangle\in\mathcal{H}_2$.
Recall that for $X$ to be an entanglement witness, the operator itself cannot be semi-positive definite.

Additionally, for the case of two qubits it can be shown \cite{Sarbicki_2008} that any entanglement witness has exactly one negative eigenvalue,  resulting in the following set of conditions for an operator to be an entanglement witness.
\medskip
\begin{prop}\label{prop:eq_main}
    An operator $X\in\mathcal{L}(\mathcal{H}_2\otimes\mathcal{H}_2)$ is an entanglement witness if and only if
    \begin{enumerate}[i)]
        \item $X$ is a Hermitian matrix;
        \item $X$ has exactly one negative eigenvalue \cite{Sarbicki_2008, Augusiak_2008}; and
        \item $\forall |w\rangle\in\mathcal{H}_2,\;\tr(X_w)\geq0\land\det(X_w)\geq0$.
    \end{enumerate}
\end{prop}
Conditions i) and ii) provide first two steps for determining if a two qubit operator is an entanglement witness. In the next two sections 
we shall provide an effective method for checking condition iii), starting with $\tr(X_w)\geq 0\;\forall w$ and then proceeding to $\det(X_w)\geq 0\;\forall w$. In both cases we assume that $X$ is a Hermitian operator.
\subsection{Non-negativity of trace of $X_w$}
 The trace condition from Proposition \ref{prop:eq_main} can be expressed explicitly in terms of elements of matrix $X$. The operator $X_w$ is Hermitian and its trace is a real number given by
\begin{eqnarray}\label{eq:trxw}
    \tr(X_w) = \left|w_1\right|^2\tau_1+\left|w_2\right|^2\tau_2+2\mathrm{Re}{\left(\overline{w_1}w_2\xi\right)},
\end{eqnarray}
where
\begin{eqnarray}\label{eq:sub_rho_tau}
    \tau_1 = X_{11,11}+X_{21,21},\quad
    \tau_2 = X_{12,12}+X_{22,22}\quad \mathrm{and}\quad
    \xi = X_{11,12}+X_{21,22}.
\end{eqnarray}
\begin{prop}\label{prop:tr_cond}
    Projected operator $X_w$ defined in (\ref{eq:defXw}) satisfies $\tr(X_w)\geq 0$ for all $|w\rangle\in\mathcal{H}_2$ if and only if
    \begin{eqnarray}\label{eq:tr_cond}
        \tr(X) = \tau_1+\tau_2\geq 0\;\land\;\tau_1\tau_2\geq\left|\xi\right|^2,
    \end{eqnarray}
    where $\tau_1$, $\tau_2$ and $\xi$ are defined in terms of matrix elements of $X$ as in (\ref{eq:sub_rho_tau}).
\begin{proof}
It is clear that the phases of the complex numbers $w_1$ and $w_2$ only enter the last term in (\ref{eq:trxw}). By choosing them properly, we can always make this term equal to $-2\left|w_1\right|\left|w_2\right|\left|\xi\right|$ and be left with the task of checking non-negativity of $\left|w_1\right|^2\tau_1+\left|w_2\right|^2\tau_2-2\left|w_1\right|\left|w_2\right|\left|\xi\right|$.
Focusing first on the case $\tau_1>0$ and $\tau_2>0$, substitute $\left|w_1\right|\rightarrow x_1/\sqrt{\tau_1}$, $\left|w_2\right|\rightarrow x_2/\sqrt{\tau_2}$ in (\ref{eq:trxw}). This leaves us with the non-negativity condition for $x_1^2+x_2^2-2x_1x_2\left|\xi\right|/\sqrt{\tau_1\tau_2}=\left(x_1+x_2\right)^2+\left(1-\left|\xi\right|/\sqrt{\tau_1\tau_2}\right)x_1x_2$ for arbitrary $x_1\geq 0$, $x_2\geq 0$, or equivalently, $\tau_1\tau_2\geq\left|\xi\right|^2$. Under our positivity assumption for $\tau_1$, $\tau_2$, this is equivalent to (\ref{eq:tr_cond}). When $\tau_1$ is zero, we get $\min\tr\left(X_w\right)=\left|w_2\right|^2\tau_2-2\left|w_1\right|\left|w_2\right|\left|\xi\right|$, which can only be non-negative for all $w_1$ and $w_2$ if $\xi=0$ and $\tau_2\geq 0$, which is again equivalent to (\ref{eq:tr_cond}). The same logic applies when we switch from $\tau_1$ to $\tau_2$. Finally, when either $\tau_i$ is negative, $\tr\left(X_w\right)$ cannot always be non-negative and conditions (\ref{eq:tr_cond}) cannot be fulfilled. Thus, non-negativity of $\tr\left(X_w\right)$ translates to conditions (\ref{eq:tr_cond}) also in this case.
\end{proof}
\end{prop}

\subsection{Non-negativity of determinant of $X_w$}\label{detsection}
To determine if $\det(X_w)\geq 0$, one can express the (complex) coordinates $w$ in terms of their real and imaginary parts and write the determinant of $X_w$ as a polynomial in these variables. The approach followed in this section is to rephrase the resulting polynomial non-negativity problem as a set of logical statements of the form
\begin{eqnarray}
    g(t)\geq0,\;\;\forall t \in(a,b)\;\;\mathrm{or}\;\;g_1(t)\geq0\;\lor\; g_2(t)\geq 0,\;\;\forall t\in(a,b),
\end{eqnarray}
where $g(t)$ are real, univariate polynomials. Truth value of such statements can be evaluated using algorithms presented in Appendix \ref{Algorithms}, which are exact for polynomials over the rationals.
If $w_1=0$ is substituted in (\ref{eq:defXw}), the condition for non-negative determinant of $X_w$ for all $w_2$ reduces to 
\begin{eqnarray}\label{eq:cos0}
    X_{12,12}X_{22,22}-\left|X_{12,22}\right|^2\geq 0.
\end{eqnarray}
Otherwise, one can substitute
\begin{eqnarray}\label{eq:subsub}
    |w\rangle = R\left(\begin{array}{c}
    1 \\
    re^{i\varphi}
    \end{array}\right),
\end{eqnarray}
where $r\in \mathbb{R}$. The normalization factor $R$ does not affect the sign of the determinant and for the purpose of this analysis can be safely set to $1$. With this substitution, the determinant becomes
\begin{eqnarray}
\fl
    \det(X_w) = 
    c_1r^4
    +\left(c_2e^{i\varphi}+\overline{c}_2e^{-i\varphi}\right)r^3
    +\left(c_3e^{2i\varphi}+\overline{c}_3e^{-2i\varphi}+c_4\right)r^2+\nonumber\\
    +\left(c_5e^{i\varphi}+\overline{c}_5e^{-i\varphi}\right)r
    +c_6, \label{eq:det_1}
\end{eqnarray}
where
\begin{eqnarray}\label{eq:abecadło}\eqalign{
        c_1 = X_{12,12}X_{22,22}-\left|X_{12,22}\right|^2 \in \mathbb{R},\\
        c_2 = X_{22,22}X_{11,12}+X_{12,12}X_{21,22}-X_{12,22}\overline{X}_{12,21}-\overline{X}_{12,22}X_{11,22} \in \mathbb{C},\\
        c_3 = X_{11,12}X_{21,22}-X_{11,22}\overline{X}_{12,21} \in \mathbb{C},\\
        c_4 = X_{11,12}\overline{X}_{21,22}-X_{12,22}\overline{X}_{11,21}+\overline{X}_{11,12}X_{21,22}-\overline{X}_{12,22}X_{11,21}+\\
        \quad\quad+X_{11,11}X_{22,22}+X_{12,12}X_{21,21} -\left|X_{11,22}\right|^2-\left|X_{12,21}\right|^2\in\mathbb{R},\\
        c_5 = X_{11,11}X_{21,22}+X_{21,21}X_{11,12}-\overline{X}_{11,21}X_{11,22}-X_{11,21}\overline{X}_{12,21} \in \mathbb{C},\\
        c_6 = X_{11,11}X_{21,21}-\left|X_{11,21}\right|^2 \in \mathbb{R}.}
\end{eqnarray}

Expression (\ref{eq:det_1}) can be rearranged into 
\begin{eqnarray}\label{eq:det_2}
        \fl\det(X_w) = c_1r^4+2\left(\mathrm{Re}{(c_2)}\cos{\varphi}-\mathrm{Im}{(c_2)}\sin{\varphi}\right)r^3+(2\mathrm{Re}{(c_3)}\left(2\cos^2{\varphi}-1\right)-\nonumber\\4\;\mathrm{Im}{(c_3)}\sin{\varphi}\cos{\varphi}+c_4)r^2
        +2\left(\mathrm{Re}{(c_5)}\cos{\varphi}-\mathrm{Im}{(c_5)}\sin{\varphi}\right)r+c_6,
\end{eqnarray}
where trigonometric functions can be expressed using the stereographic projection,
\begin{eqnarray}\label{eq:stproj}
    \cos{\varphi} = \frac{1-t^2}{1+t^2}\quad \mathrm{and} \quad\sin{\varphi} = \frac{2t}{1+t^2},
\end{eqnarray}
with $t\in\mathbb{R}$. Substituting (\ref{eq:stproj}) into (\ref{eq:det_2}) and multiplying the expression by a positive factor $(t^2+1)^2$, results in a bivariate polynomial of the form
\begin{eqnarray}\label{eq:W_polynomial}
        \fl W(r,t) =  c_1(t^2+1)^2r^4-2(t^2+1)\Lambda\left(t,c_2\right)r^3\nonumber
        +\chi\left(c_3,c_4,t\right)r^2-2(t^2+1)\Lambda\left(t,c_5\right)r\\+\;c_6(t^2+1)^2,
\end{eqnarray}
where we introduced auxiliary expressions,
\numparts
\begin{eqnarray}\label{eq:chilambda}
    \Lambda(t,c_i) = \mathrm{Re}{(c_i)}t^2+2\mathrm{Im}{(c_i)}t-\mathrm{Re}{(c_i)}\\
    \fl\chi\left(t,c_3,c_4\right) = (c_4+2\mathrm{Re}{(c_3)})t^4+8\mathrm{Im}{(c_3)}t^3+2(c_4-6\mathrm{Re}{(c_3)})t^2-8\mathrm{Im}{(c_3)}t\nonumber\\+\;(c_4+2\mathrm{Re}{(c_3)}).
\end{eqnarray}
\endnumparts
Checking if $\det(X_w)\geq0$ for all $|w\rangle\in\mathcal{H}_2$ now comes down to finding positivity conditions for
\begin{eqnarray}
    W(r,t)\geq 0, \quad \forall t\in\mathbb{R}\;\forall r\in\mathbb{R}.
\end{eqnarray}
One can easily check that $\Lambda(t,c_i) = 0$ for all $t$ happens only if $c_i=0$ and $\chi(t,c_3,c_4)=0$ for all $t$ only if both $c_3$ and $c_4$ are zero. By a  similar token to (\ref{eq:subsub}), one can instead substitute $w_2 = 1$ and $w_1=re^{-i\varphi}$ in the expression for $\det(X_w)$. Applying similar steps as before results in an expression,
\begin{eqnarray}\label{eq:V_polynomial}
        \fl V(r,t) =  c_1(t^2+1)^2-2(t^2+1)\Lambda\left(t,c_2\right)r+
        \chi\left(c_3,c_4,t\right)r^2-2(t^2+1)\Lambda\left(t,c_5\right)r^3\nonumber\\+\;c_6(t^2+1)^2r^4,
\end{eqnarray}
that is similar to (\ref{eq:W_polynomial}). Non-negativity conditions for $W(r,t)$ that ensure $\det(X_w)\geq0$ for all $t$ and $r$ also have to result in non-negative values of $V(r,t)$.

It is clear from (\ref{eq:cos0}) that the necessary requirement for non-negativity of $W(r,t)$ is $c_1\geq0$. Similarly, checking $W(r=0,t)$, one obtains that $c_6\geq0$. If $c_1$ is positive, then $W(r,t)$ is a bivariate polynomial of $4$-th degree in both $r$ and $t$. In cases for which both $c_1$ and $c_6$ are positive, our strategy is to apply the non-negativity conditions for quartic polynomials derived by Song and Zhang \cite{Polynomial_Positivity_Criteria}. Theorem 4.3 therein along with some of its implications is discussed in in Appendix \ref{Detour_on_Polynomials}. It follows from Proposition \ref{prop:inversecoeffs}, that 
whenever $W(r,t)$ satisfies these non-negativity conditions, the same conditions are also satisfied by the polynomial $V(r,t)$. Let
\begin{eqnarray}\label{eq:fW}
    f_W(t,r) = r^4+\alpha_Wr^3+\beta_Wr^2+\gamma_W r+1,
\end{eqnarray}
where
\begin{eqnarray}\label{eq:w_alpha_beta_gamma}
    \fl\quad\quad\alpha_W  = \frac{-2\Lambda(t,c_2)}{\sqrt[4]{c_1^3c_6}(t^2+1)},\quad
    \beta_W = \frac{\chi\left(t, c_3,c_4\right)}{\sqrt{c_1c_6}(t^2+1)^2}\quad\mathrm{and}\quad
    \gamma_W =  \frac{-2\Lambda(t,c_5)}{\sqrt[4]{c_1c_6^3}(t^2+1)}
\end{eqnarray}
are defined by (\ref{eq:alpha_beta_gamma}) in Appendix \ref{Detour_on_Polynomials}. According to Theorem 4.3 in \cite{Polynomial_Positivity_Criteria}, the polynomial $f_W$ is non-negative if expressions (\ref{eq:w_alpha_beta_gamma}) and discriminant $\Delta(f_W)$ satisfy
\numparts
\begin{eqnarray}\label{eq:4.3i}
    \Delta(f_W)\geq0,\\ \left|\alpha_W-\gamma_W\right|\leq4\sqrt{\beta_W+2}\label{eq:4.3ii}
\end{eqnarray}
and
\begin{eqnarray}\label{eq:4.3iii}
    -2\leq\beta_W\leq6\;\lor\;\left(\beta_W>6\;\land\;\left|\alpha_W+\gamma_W\right|\leq4\sqrt{\beta_W-2}\right).
\end{eqnarray}
\endnumparts
Condition (\ref{eq:4.3i}) on the discriminant of $f_W$ evaluates to checking if its numerator, a $16$-th degree polynomial $g_\Delta(t)$ is non-negative for all real $t$. This can be achieved with help of Sturm theorem by applying the strategy presented in Proposition \ref{prop:poly_nonnegativity_cond}. The full expression for $g_\Delta$ is provided in Appendix \ref{polynomials}.  Remaining inequalities can also be rephrased as conditions on polynomials in $t$ by plugging in expressions for $\alpha_W$, $\beta_W$, $\gamma_W$ and rearranging the inequalities. Introducing the notation
\begin{eqnarray} \label{eq:g1poly}
    g_1(t) = 4c_1c_6\chi(t,c_3,c_4+2\sqrt{c_1c_6})-\Lambda^2(t,\sqrt{c_1}c_5-\sqrt{c_6}c_2),
\end{eqnarray}
\begin{eqnarray}\label{eq:g2poly}
    g_2(t) = \chi\left(t,c_3,c_4+2\sqrt{c_1c_6}\right),
\end{eqnarray}
\begin{eqnarray}\label{eq:g3poly}
    g_3(t) = \chi\left(t,-c_3,6\sqrt{c_1c_6}-c_4\right),
\end{eqnarray}
\begin{eqnarray}\label{eq:g4poly}
    g_4(t) = 4c_1c_6\chi(t,c_3,c_4-2\sqrt{c_1c_6})-\Lambda^2(t,\sqrt{c_1}c_5+\sqrt{c_6}c_2),
\end{eqnarray}
one obtains that (\ref{eq:4.3ii}) and (\ref{eq:4.3iii}) are equivalent to
\begin{eqnarray}
    g_1(t)\geq 0
\end{eqnarray}
and
\begin{eqnarray}
g_2(t)\geq0\;\land\;\left(g_3(t)\geq0\lor g_4(t)\geq 0\right) ,
\end{eqnarray}
respectively.
The expressions for $g_1$, $g_2$, $g_3$ and $g_4$ are $4$-th degree polynomials in $t$. Consequently, non-negativity conditions for $W(r,t)$ can be summarized as follows.
\medskip
\begin{prop}\label{prop:check}
    Bivariate polynomial $W(r,t)$ from ($\ref{eq:W_polynomial}$), with $c_1>0$ and $c_6>0$ is non-negative for all $t\in\mathbb{R}$ and all $r\in\mathbb{R}$ iff for all real $t$ i) $g_\Delta(t)\geq 0$, ii) $g_1(t)\geq0$, iii) $g_2(t)\geq0$ and iv) $g_3(t)\geq0\lor g_4(t)\geq 0$.
\end{prop}
 The key notion is that all the functions $g_i$ are polynomials in $t$ and the propositions of Appendices \ref{Detour_on_Polynomials} and \ref{Algorithms} allow one to check validity of conjunctions and alternatives of their non-negativity conditions for all $t$. The detailed approach can be summarized by the following

\medskip
 
\begin{prop}\label{prop:approach}
    Conditions presented in Proposition \ref{prop:check} can be effectively checked using the following steps:
    \begin{enumerate}[1.]
        \item To check condition i) one can apply Proposition \ref{prop:poly_nonnegativity_cond} (see Appendix \ref{Detour_on_Polynomials}).
        \item To check conditions ii) and iii) one can apply Proposition \ref{prop:first_simple_case} or again Proposition~\ref{prop:poly_nonnegativity_cond}.
        \item In order to check condition iv) one can apply the algorithm described in Appendix~\ref{Algorithms}.
    \end{enumerate}
\begin{proof}
    Proposition \ref{prop:poly_nonnegativity_cond} discusses conditions satisfied by a real univariate polynomial $f$, such that $f(t)\geq0$, $\forall t\in\mathbb{R}$. Same conditions have to be satisfied by polynomials in inequalities i), ii) and iii) of Proposition \ref{prop:check}. Finally, the alternative iv) of two polynomial non-negativity statements, $g_4(t)\geq0\;\lor\;g_5(t)\geq0$ for all $t$ is covered in Proposition \ref{prop:alternative} (see Appendix \ref{Algorithms}).
\end{proof}
\end{prop}

One of the assumptions in Proposition \ref{prop:check} was that both coefficients $c_1$ and $c_6$ are positive. To consider the cases in which at least one of those coefficients vanishes, we first rewrite $W(r,t)$ as
\begin{eqnarray}\label{eq:Wa}
    W(r) = a_4 r^4+a_3 r^3+a_2r^2+a_1r+a_0,
\end{eqnarray}
where coefficients $a_i$ correspond to the polynomial coefficients depending on $t$ in (\ref{eq:W_polynomial}). It is trivial to check that $W(r,t)\geq0,\;\forall r,t$ in case all $a_i$ coefficients vanish, or if one of $a_4$, $a_2$ or $a_0$ is the only non-vanishing coefficient in (\ref{eq:Wa}) and positive. Consequently, we have

\medskip

\begin{prop}[Trivial conditions]\label{prop:trivial}
If the coefficients $c_i$ in  (\ref{eq:abecadło}) satisfy any of the following
\begin{enumerate}[a)]
    \item $c_i = 0,\; i\in\{1,2,3,4,5,6\}$;
    \item $c_1>0 \land c_i = 0,\;i\in\{2,3,4,5,6\}$;
    \item $c_6>0 \land c_i = 0,\;i\in\{1,2,3,4,5\}$; or
    \item $\forall t \in\mathbb{R},\;\chi(t,c_3,c_4)\geq 0\;\land\;c_i = 0,\;i\in\{1,2,5,6\}$,
\end{enumerate}
then $W(r,t)$ is non-negative for all real $r$ and $t$.
\begin{proof}
    Each option corresponds to a trivial case with at most one non-vanishing coefficient $a$ in expression $(\ref{eq:Wa})$. The conditions of the Proposition ensure that the respective non-vanishing coefficient is positive.
\end{proof}
\end{prop}

The only remaining scenario to consider is when either $a_0$ or $a_4$ vanish, but there is more than one non-vanishing coefficient. 
\medskip
\begin{prop}\label{prop:p3}
    Bivariate polynomial $W(r,t)$ from ($\ref{eq:W_polynomial}$), with $c_1 = 0$ and $c_6>0$ is non-negative $\forall t,s \in \mathbb{R}$ if $c_2 = 0$ and
    \begin{eqnarray}\label{eq:my}
        g_5(t) = c_6\chi(t,c_3,c_4)-\Lambda^2(t,c_5)\geq 0,\;\;\forall t.
    \end{eqnarray}
    Analogously, if $c_1>0$ and $c_6 = 0$, then $W(r,t)\geq 0\;\forall t,s$ if $c_5 = 0$ and
    \begin{eqnarray}
        g_6(t) = c_1\chi(t,c_3,c_4)-\Lambda^2(t,c_2)\geq 0,\;\;\forall t.
    \end{eqnarray}
    Both polynomials are quartic in $t$, so their non-negativity can be checked by applying Proposition \ref{prop:first_simple_case} or \ref{prop:poly_nonnegativity_cond}.
\begin{proof}
    If $c_1 = 0$ and $c_2 \neq 0$ then $W(r,t)$ is a cubic polynomial in $r$ and cannot be non-negative for all $r$. When both $c_1$ and $c_2$ are zero, $W(r,t)$ reduces to a quadratic polynomial $a_2r^2+a_1r+a_0$ with well known non-negativity condition $a_1^2\leq4a_2a_0$. Equation (\ref{eq:my}) follows from applying this inequality to (\ref{eq:W_polynomial}) with $c_1=c_2=0$. The second part of the proposition can be proven by applying the same logic to the polynomial $V(r,t)$.
\end{proof}
\end{prop}

\medskip
\begin{prop}
    The determinant of matrix $X_w$ is non-negative for all $|w\rangle\in\mathcal{H}_2$ if and only if the corresponding polynomial $W(r,t)$ satisfies conditions of Proposition~\ref{prop:check}, Proposition~\ref{prop:trivial} or Proposition \ref{prop:p3}.
\begin{proof}
    By the arguments leading up to \eqref{eq:W_polynomial}, non-negativity of $\det(X_w)$ is equivalent to non-negativity of the polynomial $W(r,t)$.
    It was already shown that if the polynomial $W(r,t)$ satisfies one of the considered propositions, then it is non-negative. In order to show that all cases are covered, consider a non-negative $W(r)$ from expression $\eqref{eq:Wa}$. If $W(r)$ is a $4$-th degree polynomial in $r$ then either $a_0>0\;\forall_t$, equivalent to $c_6>0$, or alternatively $a_0=0\;\forall_t$, which happens if and only if $c_6=0$. The former case is covered by Proposition~\ref{prop:check}. In the latter, we also need to have $a_1 = 0\;\forall_t$ (or equivalently $c_5=0$), as otherwise the corresponding polynomial $V(r,t)$ would be cubic in $r$. Note that the assumed condition $a_4 > 0\;\forall_t$ (degree $W(r)$ = $4$) is equivalent to $c_1 > 0$. Therefore, the case considered exactly matches the assumptions of Proposition \ref{prop:p3} ($c_1>0$, $c_6=0$, $c_5=5$). Finally, if $W(r)$ is not of degree $4$ for some $t$, then it is not of degree $4$ for any $t$ as $c_1=0$. We need to have $a_3=0\;\forall_t$, equivalent to $c_2=0$, as otherwise $W(r,t)$ would be cubic in $r$. If now $a_0>0\;\forall_t$ (equivalent to $c_6>0$), we have $c_1=c_2=0$, $c_6>0$ and Proposition \ref{prop:p3} applies. The remaining case $c_6=c_5=c_2=c_1=0$ is covered by case d) in Proposition \ref{prop:trivial}.
\end{proof}
\end{prop}

\subsection{Effective algorithm}\label{effectiveAlgorithm}
It is now possible to combine the propositions presented in the previous sections into the following procedure.
\medskip
\begin{prop}[Algorithm]\label{prop:procedure}
In order to verify whether a given Hermitian matrix $X$ of order $4$ is an entanglement witness, it is sufficient to perform the following steps:
\begin{enumerate}
    \item Check if $X$ is positive-semidefinite. If so, then $X$ is not an entanglement witness.
     One can check the semi-positivity of $X$ by applying Sturm theorem to its characteristic polynomial for the interval $(-\infty,0]$. The existence of negative eigenvalues implies that $X$ is not positive-semidefinite. Additionally, it follows from the results of \cite{Sarbicki_2008} that if $X$ has more than one negative eigenvalue then it is not an entanglement witness.
    \item Check the trace condition (\ref{eq:tr_cond}) from Proposition \ref{prop:tr_cond} on functions $\tau_1,\;\tau_2,\;\xi$ of the entries of $X$, as defined in (\ref{eq:sub_rho_tau}). If the condition is satisfied, proceed to step (iii). If not, then $X$ is not an entanglement witness.
    \item Construct coefficients $c_i$ from ($\ref{eq:abecadło}$). Check if $c_1\geq0$ and $c_6\geq0$. If not, then $X$ is not an entanglement witness. If the condition is satisfied, proceed to step (iv).
    \item Check if coefficients $c_i$ satisfy the trivial conditions from Proposition \ref{prop:trivial}. If so, then $X$ is an entanglement witness. If not, and $c_1 = 0\;\lor\;c_6 = 0$, check conditions of Proposition \ref{prop:p3}. If they are satisfied, then $X$ is an entanglement witness. If these conditions are not satisfied, then $X$ is not block-positive. Otherwise, that is if both $c_1$ and $c_2$ are positive, proceed to the next step.
    \item Construct the polynomials $g_\Delta$, $g_1$, $g_2$, $g_3$ and $g_4$ defined in equations (\ref{eq:g1poly})-(\ref{eq:g4poly}). Check the conditions presented in Proposition~\ref{prop:check} using the approach from Proposition~\ref{prop:approach}. If they are satisfied, then $X$ is an entanglement witness. If not, then $X$ is not an entanglement witness and this marks the end of the procedure.
\end{enumerate}
\end{prop}

In case a matrix $X$ is found to be not block-positive by the above procedure, it seems desirable to also provide a certificate of non-block-positivity, i.e. a product vector $|v_0\otimes w_0\rangle$ such that $\langle v_0\otimes w_0|X|v_0\otimes w_0\rangle<0$. This turns out to be possible with an extension of the aforementioned algorithm.
\medskip
\begin{prop}
    Let $X$ be a Hermitian matrix of order $4$ with one negative eigenvalue. If any of the steps of the algorithm from Proposition \ref{prop:procedure} returns that $X$ is not block-positive, one can recover a separable vector $|v_0\otimes w_0\rangle\in\mathcal{H}_2\otimes\mathcal{H}_2$ such that $\langle v_0\otimes w_0|X|v_0\otimes w_0\rangle < 0$.
\begin{proof}
Existence of a suitable vector follows directly from Definition \ref{defi:Block_Positivity} of block-positivity. What remains to be shown, is how to recover it at the appropriate step of the algorithm. This translates into finding a vector $|w_0\rangle = (w_1,w_2)$ such that the projected matrix $X_w$ from (\ref{eq:defXw}) has a negative determinant or trace. Once a non positive-semidefinite operator $X_w$ is obtained, finding an appropriate $|v_0\rangle$ to construct a product state is trivial. 

The solution depends on which step in Proposition~\ref{prop:procedure} returns that $X$ is not an entanglement witness. Because $X$ is already assumed not to be positive-semidefinite, step (i) of the above algorithm can be skipped. If the algorithm returns false at step (ii), one of $|w_0\rangle = (1/\sqrt{\tau_1},-\xi/\left|\xi\right|\cdot 1/\sqrt{\tau_2})$, $|w_0\rangle=\left(1,0\right)$ or $|w_0\rangle=\left(0,1\right)$ yields $\tr(X_w)<0$ (see~(\ref{eq:trxw})). 

The cases considered in the remaining steps should result in a negative value of $\det(X_w)$. Failing conditions from step (iii) means that either $c_6<0$ or $c_1<0$. If the former case is true, one can take $|w_0\rangle=(1,0)$, which corresponds to setting $r=0$ in expression (\ref{eq:W_polynomial}). In the latter case, taking $|w_0\rangle = (0,1)$ sets $r=0$ in expression (\ref{eq:V_polynomial}). In both cases, any value of $t$ will work as it does not affect the vector $|w_0\rangle$. 

Recovering the result at steps (iv) and (v) relies on Remark \ref{rem:algos} and Algorithm~\ref{alg:mesh} of Appendix \ref{Algorithms}. The basic premise is that applying this algorithm to a polynomial that is not non-negative allows for recovery of at least one point, for which the polynomial takes a negative value. If step (iv) of the procedure returns false, then either $c_1=0$ or $c_6=0$. If $c_6 = 0$, let $|w_0\rangle=(1,re^{i\varphi})$, with $\varphi$ related to $t$ by (\ref{eq:stproj}). Then run Algorithm \ref{alg:mesh} from Appendix \ref{Algorithms} on $W(r,0)$ which is a quartic polynomial in $r$. In an unlikely case if this does not return $r_0$ for which $W$ is negative, it follows that $W(r,0) = c_1r^4$, and one can repeat the process for $t=+\epsilon$ until a desired value of $r_0$ is recovered. In the case of $c_1 = 0$, letting $|w\rangle =(re^{-i\varphi},1)$ and applying the same logic to $V(r,0)$ from (\ref{eq:V_polynomial}) recovers desired values of $t_0$ and $r_0$. If a matrix $X$ passes steps (i) through (iv) and returns false at the last step, it means that it does not satisfy one of the polynomial inequalities from Proposition \ref{prop:check}. One can now run Algorithm \ref{alg:mesh} on the polynomial appearing in said inequality and obtain some $t_0$ for which it is not satisfied. When evaluating the statement $g_3(t)\geq0\lor g_4(t)\geq 0$, the procedure, described in Algorithm \ref{alg:precedence} of Appendix \ref{Algorithms}, already finds an interval in which the inequalities are not satisfied. In such cases, it is easy to add a step returning some point from this interval. Having found the desired $t_0$, one can now run Algorithm \ref{alg:mesh} on $W(r,t_0)$ which returns some $r_0$ for which $W(r_0,t_0)<0$ and construct a vector $|w_0\rangle = (1,re^{i\varphi})$. 

The final required step is finding the second part of the product state, that is a vector $|v_0\rangle$ such that $\langle v_0|X_{w0}|v_0\rangle<0$. Given that $X_{w0}$ is a matrix of dimension $2$, finding such $|v_0\rangle$ is trivial. Thus, a vector $|v_0\otimes w_0\rangle$ such that $\langle v_0\otimes w_0|X|v_0\otimes w_0\rangle<0$ can be efficiently found.
\end{proof}
\end{prop}

\subsection{Weakly optimal witnesses}
If the algorithm based on Proposition \ref{prop:procedure} returns that $X$ is block-positive,  it is natural to ask whether $X$ is a weakly optimal entanglement witness, i.e. whether it belongs to the boundary of the set of entanglement witnesses \cite{Badziag_2013}. If so, then $\langle v|X_w|v\rangle = \langle v\otimes w|X|v\otimes w\rangle = 0$ for some $|w\rangle$ and $|v\rangle$ in $\mathcal{H}_2$.
\medskip
\begin{prop}
     An entanglement witness $X\in\mathcal{L}(\mathcal{H}_2\otimes\mathcal{H}_2)$ is weakly optimal if the polynomial $W(r,t)$ defined in (\ref{eq:W_polynomial}) has at least one real root.
\begin{proof}
    The sign of $W(r,t)$ corresponds to the sign of the determinant of the projected operator $X_w$. Since $X$ is an entanglement witness, $X_w$ is positive-semidefinite for any $|w\rangle$.  Hence, if $W(r,t)$ has a real root, then there exists $|w\rangle\in\mathcal{H}_2$ such that $X_w$  has a zero eigenvalue with a corresponding eigenvector $|v\rangle$ such that $\langle v\otimes w |X|v\otimes w\rangle$ = 0.
\end{proof}
\end{prop}
For $X$ that is an entanglement witness, the expression $W(r,t)$ is always non-negative. This means that for a fixed $t$, polynomial $W(r,t)$ can only have roots in $r$ of even multiplicities. A necessary condition for a quartic polynomial to have only roots of even multiplicities is that its discriminant equals zero. However, it is possible that a polynomial has a zero discriminant and does not have any real roots. One can check whether the discriminant of $W(r,t)$ equals zero for some $t_0$. However, finding if the polynomial $W(r,t_0)$ has real roots requires specifying the value of $t_0$, which might be a root of a $16$-th degree polynomial. Hence, one is left with the following necessary but not sufficient condition for weak optimality of a witnesses.
\medskip
\begin{prop}
    A witness $X\in\mathcal{L}(\mathcal{H}_2\otimes\mathcal{H}_2)$ can only be weakly optimal if its corresponding polynomial $g_\Delta$ defined in Section \ref{detsection} has real roots. This can be checked with the help of Sturm theorem (Theorem \ref{tw:Sturm Theorem} of Appendix \ref{Detour_on_Polynomials}). Moreover, this is also a necessary condition for $X$ to be an optimal \cite{Lewenstein_2000}, optimal PPTES \cite{HaKye} or extremal \cite[Section 4]{Hansen} witness.
\begin{proof}
    Polynomial $g_\Delta$ is the numerator of the discriminant of (\ref{eq:fW}). If it has any real roots then the discriminant of $W(r,t)$ also has real roots and $W(r,t)$ might have real roots of even multiplicities.
\end{proof}
\end{prop}

For a clear exposition of the theory of entanglement witness optimality, the reader is advised to consult Section 2.4 of the lecture notes \cite{KyeLectureNotes} and the related original papers \cite{Kye2012, Sarbicki2011}.
\section{The case of qudit-qubit system}\label{beyond}
With the algorithm presented in Section \ref{effectiveAlgorithm} it is possible to check whether a given Hermitian matrix of order four can serve as an entanglement witness. It might not come as a surprise that such a result can be obtained in a two qubit setting, as for these and qubit-qutrit systems there already exists a necessary and sufficient entanglement criterion due to Peres-Horodeccy \cite{HORODECKI19961}. On the other hand, it was shown \cite{Skowronek_2016} that there is no straightforward generalization of the PPT criterion to the two-qutrit system. Furthermore, the same question for qudit-qubit systems in general remains open. Let us discuss how the approach used in Section \ref{effectiveAlgorithm} might be a starting point for an algorithm detecting entanglement witnesses in these systems. 

Potential candidates for entanglement witnesses $X\in\mathcal{L}(\mathcal{H}_d\otimes\mathcal{H}_2)$ are Hermitian matrices with $2d$ rows and columns. Recall that for $X$ to be an entanglement witness, it has to satisfy the block-positivity conditions (\ref{eq:Elementwise Block-Positivity}) or (\ref{eq:defXw}). Again, the problem can be rephrased into checking semi-positivity of a projected operator $X_w$, this time of order $d$ with elements $(X_w)_{ii'} = X_{ij,i'j'}\overline{w}_jw_{j'}$ for all $w_j$, where $|w\rangle\in\mathcal{H}_2$. As discussed before, this is the same as checking the non-negativity of principal minors of order $i\in\{1,...,d\}$ of the matrix $X_w$. The procedure described in Proposition \ref{prop:procedure} can address the first two inequalities, i.e. the trace and principal minors of order $2$. For trace, a substitution analogous to ($\ref{eq:sub_rho_tau}$)
\begin{eqnarray}\label{eq:ttr}
    \tau_1 = \sum_{1}^d X_{i1,i1},\quad
    \tau_2 = \sum_{1}^d X_{i2,i2},\quad \text{and}\quad
    \xi = \sum_{1}^d X_{i1,i2},
\end{eqnarray}
gives
\begin{eqnarray}
    \tr(X_w) = \left|w_1\right|^2\tau_1+\left|w_2\right|^2\tau_2+2\mathrm{Re}{\left(\overline{w_1}w_2\xi\right)},
\end{eqnarray}
which is the same as (\ref{eq:trxw}). Conditions for non-negativity of $\tr(X_w)$ in a qubit-qudit system then follow from Proposition \ref{prop:tr_cond}.

The sum of the principal minors of order two of $X_w$, is the sum of determinants of $d$ choose $2$ Hermitian matrices of order two of the form 
\begin{eqnarray}\label{eq:Xwlk}
    X_w^{(l,k)}=
    \begin{pmatrix}
        (X_w)_{ll} && (X_w)_{lk}\\
        (X_w)_{kl} && (X_w)_{kk}
    \end{pmatrix}.
\end{eqnarray}
The indices $l\neq k$ represent rows and columns that are not removed from a given principal minor. The condition for non-negativity of their sum reads,
\begin{eqnarray}
    \sum^d_{k>l}\det(X_w^{(l,k)})\geq 0,\;\forall |w\rangle\in\mathcal{H}_2.
\end{eqnarray}
Following the steps from Section \ref{detsection}, one can rephrase the inequality concerning determinants as
\begin{eqnarray} \label{eq:w2xn}
    W(r,t) = \sum^d_{k>l}W^{(l,k)}(r,t)\geq0,\quad\forall r,t\in\mathbb{R},
\end{eqnarray}
where $W^{(l,k)}(r,t)$ is a bivariate quartic polynomial corresponding to a matrix $X_w^{(l,k)}$ constructed according to (\ref{eq:W_polynomial}). Coefficients appearing in the expressions for each polynomial are given by formulae  
\begin{eqnarray}\label{eq:abecadło2}
\eqalign{
        c_1^{(l,k)} = X_{k2,k2}X_{l2,l2}-\left|X_{l2,k2}\right|^2 \in \mathbb{R},\\
        c_2^{(l,k)} = X_{k2,k2}X_{l1,l2}+X_{l2,l2}X_{k1,k2}-X_{l2,k2}\overline{X}_{l2,k1}-\overline{X}_{l2,k2}X_{l1,k2} \in \mathbb{C},\\
        c_3^{(l,k)} = X_{l1,l2}X_{k1,k2}-X_{l1,k2}\overline{X}_{l2,k1} \in \mathbb{C},\\
        c_4^{(l,k)} = X_{l1,l2}\overline{X}_{k1,k2}-X_{l2,k2}\overline{X}_{l1,k1}+\overline{X}_{l1,l2}X_{k1,k2}-\overline{X}_{l2,k2}X_{l1,k1}+\\\quad\quad\quad+X_{l1,l1}X_{k2,k2}+X_{l2,l2}X_{k1,k1}-\left|X_{l1,k2}\right|^2-\left|X_{l2,k1}\right|^2\in\mathbb{R},\\
        c_5^{(l,k)} = X_{l1,l1}X_{k1,k2}+X_{k1,k1}X_{l1,l2}-\overline{X}_{l1,k1}X_{l1,k2}-X_{l1,k1}\overline{X}_{l2,k1} \in \mathbb{C},\\
        c_6^{(l,k)} = X_{l1,l1}X_{k1,k1}-\left|X_{l1,k1}\right|^2 \in \mathbb{R},
        }
\end{eqnarray}
analogous to coefficients in (\ref{eq:abecadło}). Condition (\ref{eq:w2xn}) can now be tackled as described in Propositions~\ref{prop:check}~and~\ref{prop:trivial}. Since all principal minors of a positive semidefinite matrix are nonnegative, instead of condition (\ref{eq:w2xn}) one can consider $d$ choose $2$ separate inequalities of the form
\begin{eqnarray}\label{eq:lastcon}
    W^{(l,k)}(r,t) \geq 0, \quad \forall r,t \in \mathbb{R},
\end{eqnarray}
where pairs of indices $(l,k)$ are such that $k>l$. Inequalities (\ref{eq:lastcon}) provide a necessary condition for block-positivity that is possibly stronger than the one from Eq. (\ref{eq:w2xn}).

Non-negativity conditions for principal minors of order $n\geq3$ is where the algorithm is no longer applicable. Using similar methods to those presented in Section \ref{detsection}, we arrive at positivity conditions for bivariate polynomial of  degree $2n$. and solving the problem for $n\geq3$ goes beyond the scope of methods used in proving Proposition~\ref{prop:procedure}. In particular, the results of \cite{Polynomial_Positivity_Criteria}, summarized in Appendix \ref{Detour_on_Polynomials}, do not apply to polynomials of degree greater than $5$. However, the results described above hold, and can be summarized as follows.
\medskip
\begin{prop}\label{prop:beyond}
    Let $X$ be a hermitian operator in $\mathcal{L}(\mathcal{H}_d\otimes\mathcal{H}_2)$ with the corresponding polynomials $W^{(l,k)}(r,t)$ constructed from matrices (\ref{eq:Xwlk}), and values $\tau_1$, $\tau_2$ and $\xi$ defined in (\ref{eq:ttr}). If any of the polynomials $W^{(l,k)}(r,t)$ does not satisfy the conditions discussed in Propositions~\ref{prop:check}~and~\ref{prop:trivial}, or if $\tau_1$, $\tau_2$ and $\xi$ do not satisfy conditions of Proposition~\ref{prop:tr_cond}, then $X$ is neither block-positive nor an entanglement witness.
\end{prop}

\section{Gröbner basis approach}\label{s5}
An alternative approach to checking the block-positivity of $X\in\mathcal{L}(\mathcal{H}_{d_1}\otimes\mathcal{H}_{d_2})$ can be taken with help of Gröbner bases (for detailed definitions and their application see \cite{ideals}). Consider again the projected operator $X_w$ of order $d_2$ defined in (\ref{eq:defXw}), and the sums of its minors of order $n$, which we here denote as $\mathcal{M}_n(w)$. Rewriting each complex component of $|w\rangle\in\mathcal{H}_{d_1}$ in terms of its real and imaginary parts $w_j = w^\mathrm{r}_j+\mathrm{i}w^\mathrm{i}_j$ turns each $\mathcal{M}_n$ into a polynomial of $2n$-th degree in $d_1$ variables $w^\mathrm{r}$ and $d_1$ variables $w^\mathrm{i}$. Since vectors $|w\rangle$ are equivalent up to a phase, we can always reduce the number of variables by one by choosing $w^\mathrm{i}_1=0$. The problem of checking whether the polynomial $\mathcal{M}_n(w)$ is non-negative for all $|w\rangle$ (assuming it is not trivially negative) can be rephrased as checking if there exists its negative minimum. That is, if $X$ is not block-positive, then there exists a non-zero $k$ that is a root of a polynomial $\mathcal{M}_n(w)+k^2$ for some $|w\rangle$. Restricting the problem to normalized states only, introduces a constraint
\begin{eqnarray}
    F(w)=\sum_i^d|w_i^\mathrm{r}|^2+|w_i^\mathrm{i}|^2-1=0.
\end{eqnarray}
In order to find $k$ it is enough to check the extreme points of $\mathcal{M}_n(w)+k^2$, which in principle can be attempted using Lagrange multiplier method. Let
\begin{eqnarray}
    L_n(w,k,\lambda)=\mathcal{M}_n(w)+k^2+\lambda F(w)
\end{eqnarray}
be the Lagrange function for given $n$, and construct a set of $2d+1$ equations of the form
\begin{eqnarray}\label{eq:ideal}
    \partial_jL_n(w,k,\lambda)=\partial_j \left(\mathcal{M}_n(w)+k^2 +\lambda F(w)\right)=0,
\end{eqnarray}
by differentiating with respect to each variable $j$. Polynomials $\{l_1,...,l_{d_1+1}\}$ appearing in (\ref{eq:ideal}) generate an ideal $I$ (cf. \cite{ideals} for basic definitions). The system of equations can now be rephrased in terms of Gröbner basis elements $g_j$ of $I$, with the last element, denoted as $g_{-1}$, being a univariate polynomial in $k$. Such a basis can be found using algorithms presented in \cite{ideals}. With help of Sturm theorem one can then determine whether equation $g_{-1}(k) = 0$ has any real roots other than zero. If not, then $\mathcal{M}_n(w)+k^2=0$ has no real negative solutions and the sum of minors of order $n$ is non-negative. If all minors of orders $n\leq d_2$ are checked and return no solutions, then $X$ is a block-positive operator.    
\medskip
\begin{prop}\label{prop:gb}
    An operator $X\in\mathcal{L}(\mathcal{H}_{d_1}\otimes\mathcal{H}_{d_2})$ is block-positive if for each $n\in\{1,...,d_2\}$ the system of equations (\ref{eq:ideal}) has no real solutions. Otherwise, the test is inconclusive. For $n=1$ and $n=2$ one can obtain a conclusive answer using the approach of Section \ref{beyond}.
\end{prop}

\section{Exemplary families of entanglement witness}\label{s6}
\begin{figure}[h]
    \centering
    \includegraphics[width = .75\linewidth]{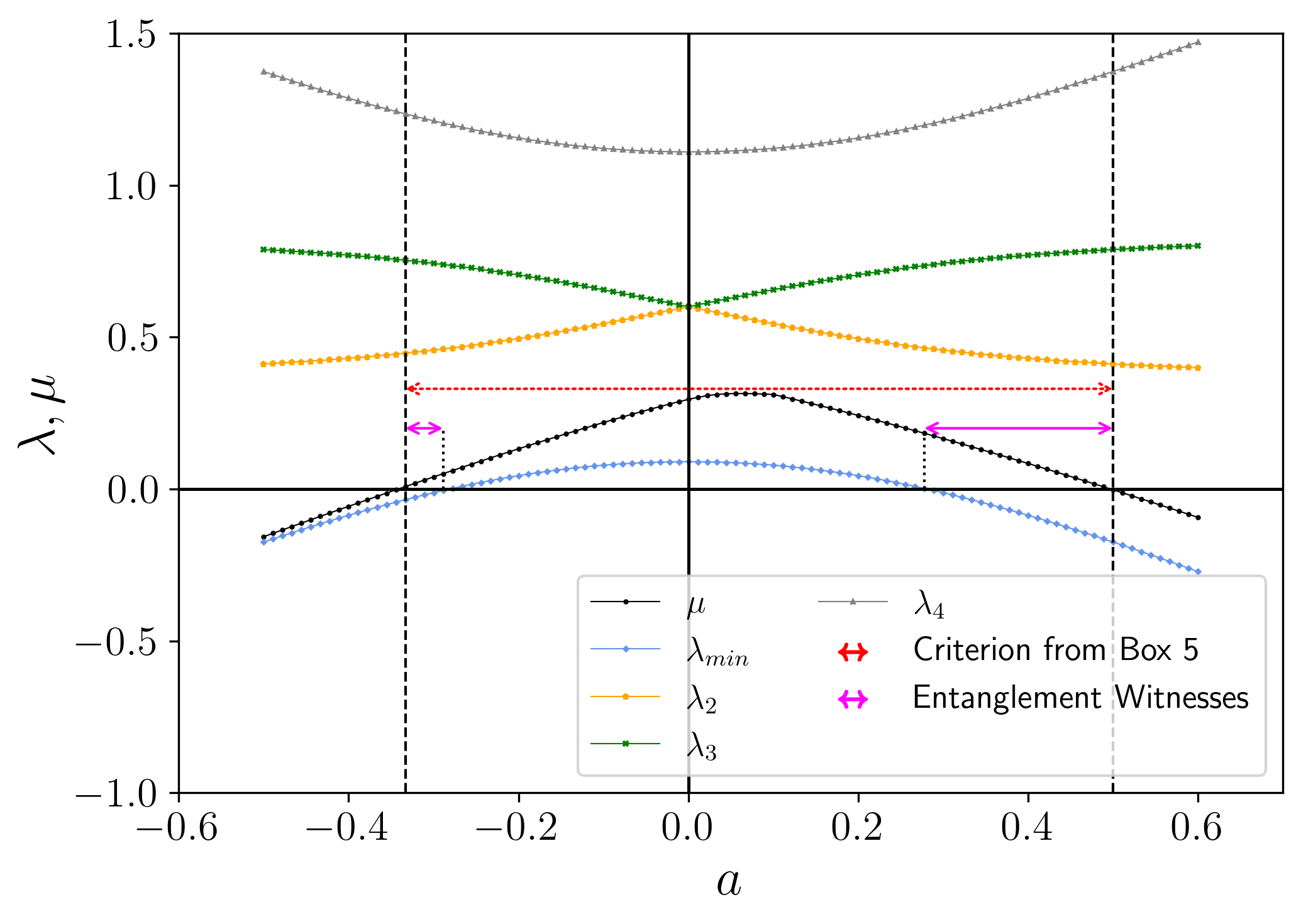}
    \caption{Eigenvalues $\lambda_i$ and minimal local value $\mu$ of operators $E[a]$ from (\ref{eq:ops}). Arrows mark intervals on which operators are block-positive (red dotted), and block-positive but not positive (pink solid), thus qualifying as entanglement witnesses.}
    \label{fig:2}
\end{figure}

\begin{figure}[h]
    \centering
    \includegraphics[width=0.75\linewidth]{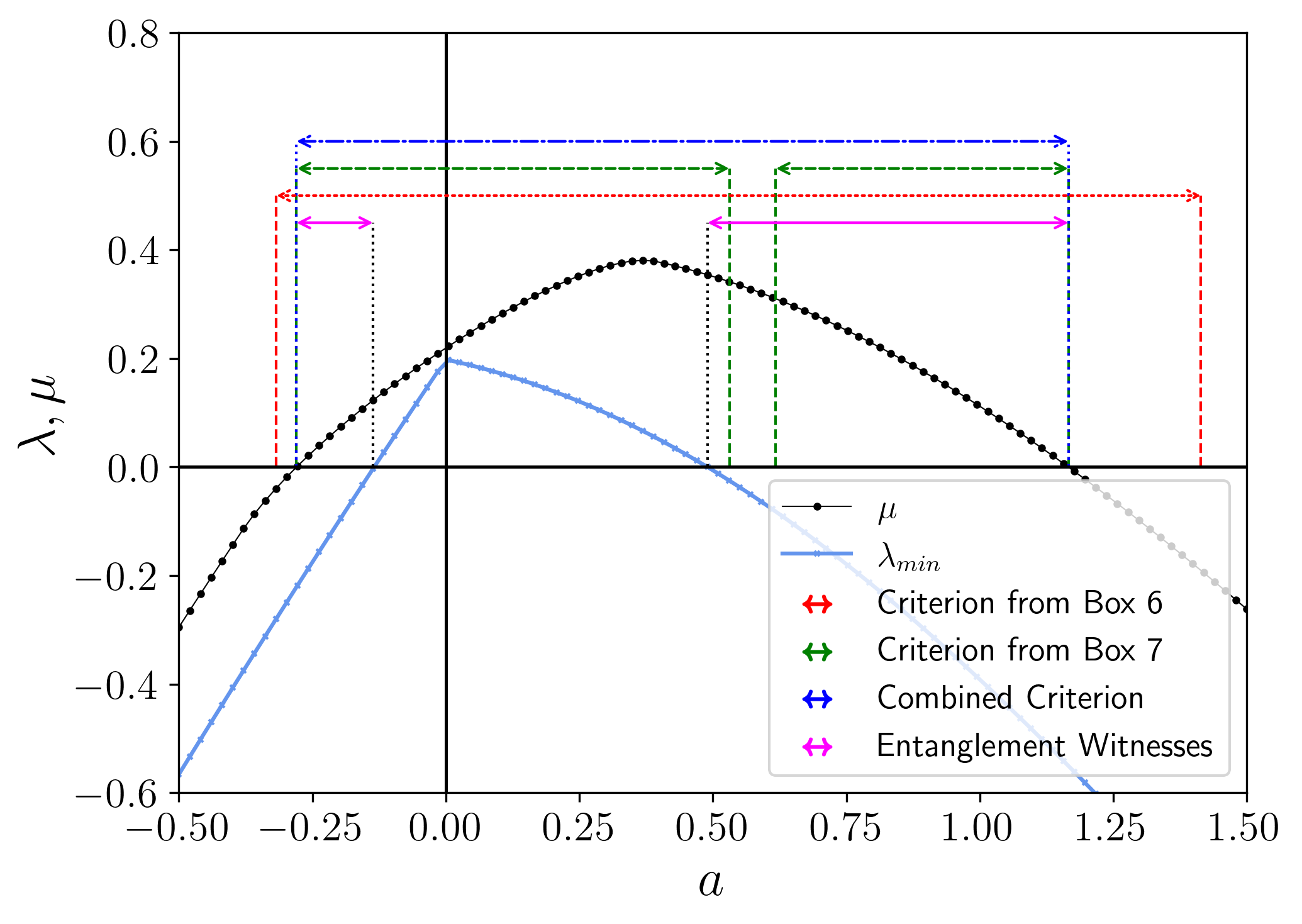}
    \caption{Minimum eigenvalue $\lambda_{min}$ and minimal local value $\mu$ of operators $F[a]$ from (\ref{eq:ops}). Arrows mark the intervals on which the different block-positivity criteria are satisfied.}
    \label{fig:3}
\end{figure}

It this section we demonstrate a practical application of the techniques developed above to identify block-positive matrices and entanglement witnesses.
Consider the following one parameter families of matrices of order $4$ and $6$ respectively,
\begin{eqnarray}\label{eq:ops}
\fl E[a]=
\begin{pmatrix}
    \frac{3}{5} & \frac{1}{10} & 0 & \frac{a}{2}(1+i) \\  \frac{1}{10}& \frac{3}{5} & -\frac{1}{2} & 0 \\
    0 & -\frac{1}{2} & \frac{3}{5} & -a \\
    \frac{a}{2}(1-i) & 0 & -a & \frac{3}{5}
\end{pmatrix},\;
F[a] =
\begin{pmatrix}
    2+a & -a & 0 & 1 & 1 & 0 \\
    -a & 1+a & 0 & 0 & 0 & 1 \\
    0 & 0 & 2 & a & a & 1 \\
    1 & 0 & a & 2 & 0 & 0 \\
    1 & 0 & a & 0 & 1 & 0 \\
    0 & 1 & 1 & 0 & 0 & 2+a
\end{pmatrix}.
\end{eqnarray}
Let $\mu(X)$ denote the value defined in (\ref{eq:mu}), that is the minimum expectation value of the operator $X$ acting on product states. Figures \ref{fig:2} and \ref{fig:3} illustrate how a numerical estimate of $\mu$ varies with the parameter $a$ for operators $E[a]$ and $F[a]$ respectively. Additionally, Figure~\ref{fig:2} includes all eigenvalues $\lambda_i$ of $E[a]$. Horizontal solid arrows indicate on which intervals the family $E$ (numerically) gives entanglement witnesses, that is the intervals for which $\mu>0$ but $\lambda_\mathrm{min}<0$. The dotted arrow marks the interval on which the criterion for block positivity from Proposition~\ref{prop:procedure} returns a positive answer (here we ignore the first step for checking positivity). Similarly, Figure \ref{fig:3} shows the lowest eigenvalue of $F[a]$, with intervals for which the criteria from Propositions~\ref{prop:beyond}~and~\ref{prop:gb} return true. Here, the necessary criterion (red dotted arrow) is satisfied for a broader set of operators, whereas criterion~\ref{prop:gb} (green dashed arrows) does not detect entanglement witnesses in the middle of the graph. However, if a combined criterion, that is minors of order $1$ and $2$ of $F_w$ are checked according to Proposition \ref{prop:beyond} and Gröbner bases are used for the determinant of $F_w$, one can detect all entanglement witnesses from the family$F[a]$.

\section{Concluding Remarks}

In this work, we have presented an effective procedure for testing whether a given Hermitian matrix $X$ forms an entanglement witness in two-qubit systems. This method is numerically exact if the coefficients of matrix $X$ are rationals extended by $i$. Our approach is based on Sturm sequences for univariate real polynomials and on positivity criteria for quartic polynomials presented in a recent paper \cite{Polynomial_Positivity_Criteria}. The problem is translated to checking positivity of a bivariate real polynomial of degree $4$, which allowed us to provide an exact procedure, described in Proposition \ref{prop:procedure}. 

Our main result should not be confused with the well-know positive partial transpose (PPT) criterion \cite{HORODECKI19961} for two qubit entanglement, which also applies to four dimensional systems, but answers a different set membership question. Here, we were interested in discriminating members of the dual set of separable states from non-members, and not in separability itself. To the best of our knowledge, there does not seem to exist an immediate way of solving these two membership problems as one, though they are by nature very closely related. 

We also touched upon extensions of the current approach beyond the two-qubit case, where additionaly, Gröbner basis methods \cite{ideals} from algebraic geometry can be applied. In this more general case we are able to provide sufficient but not necessary criteria, relating to the fact that there is no simple rule concerning the extension of partial solutions to real algebraic equations to more variables.
\section*{Acknowledgements}
\begin{sloppy}
We are thankful to Seung-Hyeok Kye for several constructive remarks that helped us to
 improve this work.
 Discussions with Ingemar Bengtsson and Markus Grassl were helpful in deepening the authors' insights into the topics covered in Section~\ref{s5}. We acknowledge funding from the EU's research programme Horizon Europe under the ERC Advanced Grant TAtypic, project number 101142236, the DQUANT QuantERA II project no. 2021/03/Y/ST2/00193 and research grant Sonata BIS 2023/50/E/ST2/00472 financed by the National Science Centre in Poland.
\end{sloppy}

\section*{References}
\bibliographystyle{iopart-num}
\bibliography{Bibliography}
\appendix

\renewcommand{\thesection}{\Alph{section}}
\setcounter{section}{0}

\section{Conditions for non-negativity of polynomials \label{Detour_on_Polynomials}}

Let $f$ denote a univariate real polynomial of $n$-th degree expressed as \cite{Tura, Basu},
\begin{eqnarray}\label{eq:generic_polynomial}
    f(t) = \sum_{i=0}^n a_it^i = g(t)\prod_i^d(t-\lambda_i)^{n_i},
\end{eqnarray}
where $g$ is a polynomial with no real roots, $d$ is a number of real roots of $f$ and $n_i$ denotes multiplicity of $i$-th real root $\lambda_i$. In order to find the greatest common divisor (GCD) of any two real polynomials $f_1,f_0$, denoted as $\mathrm{gcd}(f_1,f_0)$ one can apply the well-known Euclid's algorithm \cite{Tura}.
Applying Euclid's algorithm to $f$ from (\ref{eq:generic_polynomial}) and its derivative $f'$ returns
    \begin{eqnarray}\label{eq:GCD}
        \textrm{gcd}(f,f') = p(t)\prod^d_i(t-\lambda_i)^{n_i-1},
    \end{eqnarray}
where $p$ is some polynomial with no real roots. An important feature of the above GCD is that it has all multiple roots of $f$ with their multiplicities reduced by one. Hence, one can construct a polynomial
\begin{eqnarray}\label{eq:singleroot}
    \Tilde{f}(t) = f(t)/\mathrm{gcd}(f(t),f'(t)) = \Tilde{g}(t)\prod^d_i(t-\lambda_i),
\end{eqnarray}
which has all roots of $f$ reduced to a single multiplicity.
\medskip
\begin{defi}[Sturm sequence \cite{Tura,Basu}] \label{defi:sturm_sequence}
Let $f_0$ be a polynomial with only single roots and let $f_1$ be its derivative $f_0'$. A \textit{Sturm sequence} is a chain of polynomials
\begin{eqnarray}
    S=\{f_0(t),f_1(t),-f_2(t),-f_3(t),f_4(t),f_5(t),-f_6(t),-f_7(t),...\},
\end{eqnarray}
where $f_{i+1}$ is the remainder from division of $f_{i-1}$ by $f_{i}$.  The last element of the sequence is $f_n$, with $n$ being at most the degree of polynomial $f_0$.
\end{defi}
\medskip
\begin{thm}[Sturm theorem \cite{Tura,Basu}]\label{tw:Sturm Theorem}
    Let $f$ be a real, univariate polynomial with only single roots, and let $S$ be its Sturm sequence. Let $N(t)$ denote the number of variations, i.e. number of sign changes in the sequence $S$ evaluated at a point $t$. For any real numbers $a,$ $b,$ such that $a<b$, $f(a)\neq0$ and $f(b)\neq0$, the number of roots of $f$ in the interval $(a,b)$ is equal to $N(a)-N(b)$.
\end{thm}
Non-negativity and positivity conditions for polynomials of degree up to $5$ were derived and concisely summarized in \cite{Polynomial_Positivity_Criteria}. 
Of particular interest are the non-negativity conditions for quartic polynomials
\begin{eqnarray}\label{eq:4th-poly}
    g(t) = a_4t^4+a_3t^3+a_2t^2+a_1t+a_0
\end{eqnarray}
for all real $t$. Define
\begin{eqnarray}\label{eq:sub-4th-poly}
    f(t) = t^4+\alpha t^3+\beta t^2+\gamma t + 1,
\end{eqnarray}
with coefficients
\begin{eqnarray}\label{eq:alpha_beta_gamma}
    \alpha = a_3a_4^{-\frac{3}{4}}a_0^{-\frac{1}{4}},\quad
    \beta = a_2a_4^{-\frac{1}{2}}a_0^{-\frac{1}{2}},\quad\mathrm{and}\quad
    \gamma = a_1a_4^{-\frac{1}{4}}a_0^{-\frac{3}{4}}.
\end{eqnarray}
The discriminant of such a polynomial $f$ is given as \cite{Urlich_Watson},
\begin{eqnarray}\label{eq:disc}
     \Delta(f) = 4[\beta^2-3\alpha \gamma+12]^3-[72\beta+9\alpha \beta \gamma - 2\beta^3 -27 \alpha^2 - 27 \gamma^2]^2.
\end{eqnarray}
\begin{thm}[Theorem 4.3 in \cite{Polynomial_Positivity_Criteria}] \label{tw:4.3} Suppose that $g(t)$ is a quartic polynomial with real coefficients as in (\ref{eq:4th-poly}) such that $a_4>0$ and $a_0>0$. Define a corresponding polynomial $f(t)$ as in (\ref{eq:sub-4th-poly}),  (\ref{eq:disc}) and (\ref{eq:alpha_beta_gamma}). Then $g(t)\geq 0$ and $f(t) \geq 0$ for all real $t$ iff $\Delta(f)\geq 0$, $|\alpha - \gamma|\leq 4\sqrt{\beta+2}$ and either
\begin{enumerate}[i)]
    \item $-2\leq\beta\leq6$; or
    \item $\beta > 6$ and $|\alpha + \gamma| \leq 4\sqrt{\beta-2}$.
\end{enumerate}
\end{thm}
Based on Theorem \ref{tw:4.3}, we observe the following.
\medskip
\begin{prop}\label{prop:inversecoeffs}
    Suppose that $g_1(t)=a_4 t^4+a_3t^3+a_2t^2+a_1t+a_0$ is a quartic polynomial that satisfies the non-negativity conditions presented in Theorem \ref{tw:4.3}. A quartic polynomial of the form
    \begin{eqnarray}
        g_2(t) = a_0t^4+a_1t^3+a_2t^2+a_3t+a_4,
    \end{eqnarray}
    constructed from coefficients of $g_1$ also satisfies these conditions.
    \begin{proof}
        Constructing $\alpha$, $\beta$ and $\gamma$ coefficients for $g_1$ and $g_2$ according to (\ref{eq:alpha_beta_gamma}), results in $\alpha_2 = \gamma_1$, $\beta_2 = \beta_1$ and $\gamma_2 = \alpha_1$. Plugging these into the inequalities from Theorem \ref{tw:4.3} one obtains the same non-negativity conditions as for the polynomial $g_1$, which are assumed to be satisfied.
    \end{proof}
\end{prop}
\medskip
\begin{prop}\label{prop:first_simple_case}
    Suppose that $g(t)$ is a quartic polynomial satisfying
    \begin{eqnarray}
        a_3=-a_1 \quad\text{and}\quad a_4 = a_0>0,
    \end{eqnarray}
    that is $g(t) = a_4t^4+a_3t^3+a_2t^2-a_3t+a_4$ and $f(t) = t^4+\alpha t^3+\beta t^2-\alpha t+1$. It follows that conditions in Theorem \ref{tw:4.3} simplify to
    \begin{eqnarray}
        \left|\alpha\right|\leq2\sqrt{\beta+2}\;\land\;-2\leq\beta.
    \end{eqnarray}
    as equivalent to $f\left(t\right)$ for all $t$ real.
\begin{proof}
    Assuming that $g(t)$ satisfies the requirements considered in the proposition, one obtains the discriminant
    \begin{eqnarray}
        \Delta(f)=27\frac{(4a_3^3-4a_0a_2-8a_0^2)^2((2a_0-a_2)^2+4a_3^2)}{a_0^6},
    \end{eqnarray}
    which is always non-negative. The condition $|\alpha-\gamma|\leq 4\sqrt{\beta+2}$ takes the asserted form $|\alpha|\le 2\sqrt{\beta+2}$. Additionally, $|\alpha + \gamma| \leq 4\sqrt{\beta-2}$ simplifies to $0 \leq 4\sqrt{\beta-2}$ and is always satisfied when $\beta>6$.
\end{proof}
\end{prop}
For polynomials of degree higher than $4$, one can determine if $f(t)\geq0$ for a given interval $(a,b)$ with the help of Sturm sequences. The first and obvious requirement is that the polynomial has to be non-negative at the ends of the interval i.e. $f(a)\geq0$ and $f(b)\geq0$. One can find the number of roots $d_1$ of a polynomial $f(t)$ on a given interval, by applying Sturm theorem to the corresponding $\Tilde{f}(t)$ defined in (\ref{eq:singleroot}). If $d_1=0$ then $f(t)\geq0$ for $t\in(a,b)$. Otherwise, the question comes down to finding multiplicities of roots of $f(t)$ in $(a,b)$. The polynomial $f$ satisfies $f(t)\geq0$ for all $t$, if all its roots in $(a,b)$ are of even multiplicity. By setting $f_2 = \mathrm{gcd}(f,f')$ one can again find the number of its roots $d_2$, by applying Sturm theorem to $\Tilde{f_2}$. The polynomial $f_2$ has only multiple roots of $f$, with their multiplicities reduced by one. Hence, $d_2$ is the number of multiple roots of $f$. Repeating the procedure for $f_k = \mathrm{gcd}(f_{k-1},{f'_{k-1}})$ until $d_k = 0$, results in a  sequence of numbers 
\begin{eqnarray}\label{eq:seq}
    d=\{d_1,d_2,...\}.
\end{eqnarray}
\medskip
\begin{prop}\label{prop:poly_nonnegativity_cond}
    Suppose that $f$ is a real, univariate polynomial and $d$ is a $p$-element sequence constructed as in (\ref{eq:seq}). Then $f(t)\geq0$ for $t\in(a,b)$ if and only if $f(a)\geq0$, $f(b)\geq0$ and for all even $k\leq p$, the elements of the sequence $d$ satisfy $d_k=d_{k-1}$. In~particular, $f(t)\geq 0$ for all $t\in\mathbb{R}$ iff the degree $n$ of $f$ is even, the leading coefficient $a_n>0$ and $d_k=d_{k-1}$ for all even $k\leq p$.
\end{prop}
An element $d_i$ of the sequence $d$ corresponds to the number of roots of polynomials $f_i$ and $\Tilde{f}_i$, with the latter having only roots of single multiplicity. If $d_k=d_{k+1}$ then $\Tilde{f}_k$ and $\Tilde{f}_{k+1}$ have the same roots. If $d_k>d_{k+1}$, then the polynomial $f$ has some roots of multiplicity $k$, that are single roots of $f_{k}$ and no longer appear in $f_{k+1}$. Hence, one can construct polynomials
\begin{eqnarray}
    \delta \left(f_k\right) = \frac{\Tilde{f}_{k}}{\Tilde{f}_{k+1}},\quad k = 1,2,...,
\end{eqnarray}
with only single roots, corresponding to the roots of $f$ with multiplicity $k$. This implies that another polynomial of the form
\begin{eqnarray}\label{eq:poly_sigma}
    \sigma_f = \prod_{k=0} \delta f_{2k+1},
\end{eqnarray}
comprises all odd multiplicity roots of $f$. Polynomial (\ref{eq:poly_sigma}) should have only single roots and hence can be checked by a single run of Sturm theorem.

\section{Logical statements on polynomials \label{Algorithms}}

Being able to evaluate statements such as $f(t)\geq0,\;\forall t\in(a,b)$,
allows one to check more complicated logical formulas of the form
\begin{eqnarray}
    \Theta(g_1(t)\geq0,g_2(t)\geq0,...,g_n(t)\geq 0), \; t\in(a,b),
\end{eqnarray}
where $g_i$ are univariate real polynomials. Operations such as negation
\begin{eqnarray}
    \neg (g(t)\geq0),\quad t\in(a,b) \; \Leftrightarrow \; -g(t)>0,\quad t\in(a,b)
\end{eqnarray}
and conjunction
\begin{eqnarray}
    g_1(t)\geq0\;\land\;g_2(t)\geq0,\quad t\in(a,b)
\end{eqnarray}
are easy to check as the former requires only a single application of Sturm theorem and the latter comes down to checking individual statements separately. To evaluate the alternative
\begin{eqnarray}\label{eq:alternative}
    g_1(t)\geq0\;\lor\;g_2(t)\geq0,\quad t\in(a,b),
\end{eqnarray}
one needs to apply a slightly different approach.

First, assume that the interval $(a,b)$ is divided into smaller intervals by a set of $l$ ordered points, $t_1<t_2<...<t_l$, such that in each interval $(t_{i},t_{i+1})$ there is at most one sign flip for each polynomial $g$. The method for finding a suitable set of such points is presented in Algorithm \ref{alg:mesh}.
\begin{rem}\label{rem:algos}
   The same procedure works for a single polynomial. It suffices to set the $g_1$ in Algorithm \ref{alg:mesh} to the polynomial in question and $g_2$ to $1$.
\end{rem}
It is possible to evaluate statement (\ref{eq:alternative}) by checking the signs of $g_1$ and $g_2$ at the ends of each interval $(t_i,t_{i+1})$. All possible combinations of endpoint signs on an interval are presented in Table \ref{Tab:sign_changes}. In cases 1-7, either $\mathrm{sgn}(g_1(t_i))=\mathrm{sgn}(g_1(t_{i+1}))$ or $\mathrm{sgn}(g_2(t_i))=\mathrm{sgn}(g_2(t_{i+1}))$ are positive. This means that the interval is valid and the procedure moves on to checking the next interval. Cases 10-16 are not valid and condition (\ref{eq:alternative}) doesn't hold for them, as for at least one of the boundaries both $g_1$ and $g_2$ are negative. 

In cases 8 and 9 the outcome depends on which of the polynomials changes its sign first, so they require more investigation. In order to answer the validity of \eqref{eq:alternative} one can start with finding the number of roots of the polynomial defined by $\mathrm{gcd}(\sigma_{g_1},\sigma_{g_2})$
in the interval $(t_i,t_{i+1})$, where polynomials $\sigma_{g_i}$ are defined as in (\ref{eq:poly_sigma}). This task can be accomplished using Sturm theorem. If the outcome is one, then $g_1$ and $g_2$ have a common root and the flip occurs at the same spot.
In such a case the interval is valid. The only outcome left to consider is zero, in which case $g_1$ and $g_2$ have no common roots in $(t_i,t_{i+1})$. Then one can determine the precedence of the sign flips using a bisection method as illustrated in Algorithm \ref{alg:precedence}. In case 8 from Table \ref{Tab:sign_changes}, the precedence in favor of $g_2$ implies a valid interval, while precedence in favor of $g_1$ implies that there exists $t\in(t_i,t_{i+1})$ in which both $g_1$ and $g_2$ are negative. One can apply a similar logic in order to investigate the outcomes of case 9.
\medskip
\begin{prop}\label{prop:alternative}
Consider two polynomials $g_1$ and $g_2$. One can evaluate the truth value of a statement (\ref{eq:alternative}) that for any $t$ in the interval $(a,b)$ either $g_1(t)$ or $g_2(t)$ is non-negative by applying algorithms \ref{alg:mesh} and \ref{alg:precedence} and consulting Table \ref{Tab:sign_changes}.
\end{prop}

\begin{table}[H]
\caption{\label{Tab:sign_changes}Possible sign combinations of polynomials $g_1$ and $g_2$ at the endpoints of the interval $(t_i,t_{i+1})$, when there is at most one sign change for each polynomial.}
\begin{indented}
\item[]\begin{tabular}{@{}ccccccccccccccccc}
\br
\textrm{Case:} & 1 & 2 & 3 & 4 & 5 & 6 & 7 & \textbf{8} & \textbf{9} & 10 & 11 & 12 & 13 & 14 & 15 & 16 \\
\mr
$g_1(t_i)$     & \(+\) & \(+\) & \(+\) & \(+\) & \(-\) & \(-\) & \(+\) & \(+\) & \(-\) & \(+\) & \(-\) & \(+\) & \(-\) & \(-\) & \(-\) & \(-\) \\
$g_2(t_i)$     & \(+\) & \(+\) & \(+\) & \(-\) & \(+\) & \(+\) & \(-\) & \(-\) & \(+\) & \(+\) & \(-\) & \(-\) & \(+\) & \(-\) & \(-\) & \(-\) \\
$g_1(t_{i+1})$ & \(+\) & \(+\) & \(-\) & \(+\) & \(+\) & \(-\) & \(+\) & \(-\) & \(+\) & \(-\) & \(+\) & \(-\) & \(-\) & \(+\) & \(-\) & \(-\) \\
$g_2(t_{i+1})$ & \(+\) & \(-\) & \(+\) & \(+\) & \(+\) & \(+\) & \(-\) & \(+\) & \(-\) & \(-\) & \(+\) & \(-\) & \(-\) & \(-\) & \(+\) & \(-\) \\
\br
\end{tabular}
\end{indented}
\end{table}
\section{Polynomial  $g_\Delta$}\label{polynomials}

The expression for the $16$-th degree polynomial $g_\Delta$ used in Proposition \ref{prop:procedure} reads
\begin{eqnarray}
\eqalign{\fl
    g_\Delta(t) =  -27c_6^2(1 + t^2)^4\Lambda(t, c_2)^4-2(1+t^2)^2\Lambda^3(t,c_2)\Lambda(t,c_5)(8\Lambda^2(t,c_5)-9c_6\chi(t,c_3,c_4))\\
    \fl\quad+\Lambda^2(t,c_2)(-\Lambda^2(t,c_5)(6c_1c_6(1+t^2)^4-\chi^2(t,c_3,c_4))+c_6\chi(t,c_3,c_4)(36c_1c_6(1+t^2)^4\\
    \fl\quad-\chi^2(t,c_3,c_4)))+c_1(-27c_1(1+t^2)^4\Lambda^4(t,c_5)^4+\Lambda^2(t,c_5)\chi(t,c_3,c_4)(36c_1c_6(1+t^2)^4\\-\chi^2(t,c_3,c_4)))
    +c_1(c_6(-4c_1c_6(1 + t^2)^4+\chi^2(t,c_3,c_4))^2)\\
    -2c_1(1+t^2)^2\Lambda(t,c_2)\Lambda(t,c_5)(-9\Lambda^2(t,c_5)^2\chi(t,c_3,c_4)\\+2c_6(12c_1c_6(1+t^2)^4+5\chi^2(t,c_3,c_4))).}
\end{eqnarray}
\clearpage
\section{Algortihms}

\begin{algorithm}
\caption{Pseudocode checking the precedence of a sign change between two polynomials $g_1$ and $g_2$ on the interval $(a,b)$.}\label{alg:precedence}
\textbf{Input:}
\begin{itemize}
    \item $g_1,g_2$ - real, nonzero, univariate polynomials having no common roots on interval $(a,b)$.
    \item $a,b$ - real numbers, such that $a<b$ and sings of $g_1$ and $g_2$ at points $a$, $b$ satisfy either case 8 or case 9 from Table \ref{Tab:sign_changes}.
\end{itemize}
\textbf{Output:} 
\begin{itemize}
    \item True/False - truth value of the statement: sign change of $g_1$ precedes the sign change of $g_2$.
\end{itemize}
\begin{mmaCell}{Input}

precedence[g1_, g2_, var_, a_, b_] := 
Module[\{s1, s2, mid\},
(*Bisection of the interval (a,b)*)
  mid = (a + b)/2;
(*s1 = g1(mid)g1(a), s2 = g2(mid)g2(a)
variables s1,s2 are positive as long as there is no sign change
between a and mid*)
  s1 = (g1 /. var -> mid)*(g1 /. var -> a);
  s2 = (g2 /. var -> mid)*(g2 /. var -> a);
(*sign of g1 flips before sign of g2*)
  If[And[s1 <= 0, s2 > 0], Return[True]];
(*sign of g2 flips before sign of g1*)
  If[And[s1 > 0, s2 <= 0], Return[False]];
(*Inconclusive - neither g1 nor g2 flipped on interval (a,mid)
-> apply bisection to the interval (mid,b)*)
  If[And[s1 > 0, s2 > 0], 
   Return[PolynomialPrecedence[g1, g2, var, mid, b]]];
(*Inconclusive - both g1 and g2 flipped on interval (a,mid)
-> apply bisection to the interval (a,mid)*)
  PolynomialPrecedence[g1, g2, var, a, mid]]
  
\end{mmaCell}
\end{algorithm}
\begin{algorithm}[H]
\caption{Pseudocode returning a mesh of $l$ points $\{t_1,...,t_l\}$, such that on each interval $(t_i,t_{i+1})$ polynomials $g_1$ and $g_2$ have at most one root.}\label{alg:mesh}
\textbf{Input:}
\begin{itemize}
    \item $g_1,g_2$ - real, nonzero, univariate polynomials.
    \item $a,b$ - real numbers, such that $a<b$ and $g_1$ and $g_2$ have no roots on $(-\infty,a]$ and $[b,+\infty)$.
\end{itemize}
\textbf{Output:} 
\begin{itemize}
    \item $\{t_1,t_2,...,t_l\}$ - mesh of points, such that $g_1$, $g_2$ have no roots for $t<t_1$ or $t>t_l$, none of $t_i$'s is a root of $g_1$ or $g_2$, and on each interval $(t_i,t_{i+1})$ there is at most one root of $g_1$ or $g_2$.
\end{itemize}

\begin{mmaCell}{Input}
generateMesh[g1_, g2_, var_, a_, b_] := 
Module[\{rootsG1, rootsG2, mid, leftMesh, rightMesh,
            reducedG1, reducedG2, offset\},
(*Function reduceRootMultiplicities returns a polynomial\\ with reduced root multiplicities*)
  reducedG1 = reduceRootMultiplicities[g1, var];
  reducedG2 = reduceRootMultiplicities[g2, var];
(*Function countRealRootsInInterval returns the number\\ of roots of a polynomial on (a,b)*)
  rootsG1 = countRealRootsInInterval[reducedG1, var, a, b];
  rootsG2 = countRealRootsInInterval[reducedG2, var, a, b];
(*If both polynomials have at most one root in the interval,
return the endpoints*)
  If[rootsG1 <= 1 && rootsG2 <= 1,
   	Return[\{a, b\}]];
(*Otherwise, bisect the interval*)
  mid = (a + b)/2; 
(*If mid is a root of either g1 or g2, offset mid*)
(*Function IsRoot returns T/F*)
  While[Or[IsRoot[g1, var, mid], IsRoot[g2, var, mid]],
   	offset = (mid + b)/2;
   	mid += offset];
(*Recursively generate meshes\\ for the left and right subintervals*)
  leftMesh = GenerateMesh[g1, g2, var, a, mid];
  rightMesh = GenerateMesh[g1, g2, var, mid, b];
(*Combine the meshes, removing the duplicate midpoint*)
  Return[Join[Most[leftMesh], rightMesh]]]
  
\end{mmaCell}
\end{algorithm}

\end{document}